\documentclass[aps,prl,reprint,superscriptaddress]{revtex4-1}

\usepackage{graphicx}
\usepackage{dcolumn}
\usepackage{bm}
\usepackage{amsfonts}
\usepackage{amssymb}
\usepackage{amsmath}
\usepackage{graphicx}
\usepackage{color}
\usepackage{array}
\usepackage{dcolumn} 
\usepackage{bm} 
\usepackage{amsthm}
\usepackage{latexsym}
\usepackage{float}
\usepackage{tabularx}

\begin{document}


\title{Observation of Fermi-Pasta-Ulam Recurrence Induced by Breather Solitons in an Optical Microresonator}


\author{Chengying Bao}
\email{bao33@purdue.edu}
\affiliation{School of Electrical and Computer Engineering, Purdue University, 465 Northwestern Avenue, West Lafayette, IN 47907-2035, USA}

\author{Jose A. Jaramillo-Villegas}
\affiliation{School of Electrical and Computer Engineering, Purdue University, 465 Northwestern Avenue, West Lafayette, IN 47907-2035, USA}
\affiliation{Facultad de Ingenier\'{i}as, Universidad Tecnol\'{o}gica de Pereira, Pereira, RI 66003, Colombia}

\author{Yi Xuan}
\affiliation{School of Electrical and Computer Engineering, Purdue University, 465 Northwestern Avenue, West Lafayette, IN 47907-2035, USA}
\affiliation{Birck Nanotechnology Center, Purdue University, 1205 West State Street, West Lafayette, Indiana 47907, USA}

\author{Daniel E. Leaird}
\affiliation{School of Electrical and Computer Engineering, Purdue University, 465 Northwestern Avenue, West Lafayette, IN 47907-2035, USA}

\author{Minghao Qi}
\affiliation{School of Electrical and Computer Engineering, Purdue University, 465 Northwestern Avenue, West Lafayette, IN 47907-2035, USA}
\affiliation{Birck Nanotechnology Center, Purdue University, 1205 West State Street, West Lafayette, Indiana 47907, USA}

\author{Andrew M. Weiner}
\email{amw@purdue.edu}
\affiliation{School of Electrical and Computer Engineering, Purdue University, 465 Northwestern Avenue, West Lafayette, IN 47907-2035, USA}
\affiliation{Birck Nanotechnology Center, Purdue University, 1205 West State Street, West Lafayette, Indiana 47907, USA}
\affiliation{Purdue Quantum Center, Purdue University, 1205 West State Street, West Lafayette, IN 47907, USA}
\begin{abstract}
We present, experimentally and numerically, the observation of Fermi-Pasta-Ulam recurrence induced by breather solitons in a high-Q SiN microresonator. Breather solitons can be excited by increasing the pump power at a relatively small pump phase detuning in microresonators. Out of phase power evolution is observed for groups of comb lines around the center of the spectrum compared to groups of lines in the spectral wings. The evolution of the power spectrum is not symmetric with respect to the spectrum center. Numerical simulations based on the generalized Lugiato-Lefever equation are in good agreement with the experimental results and unveil the role of stimulated Raman scattering in the symmetry breaking of the power spectrum evolution. Our results shows that optical microresonators can be exploited as a powerful platform for the exploration of soliton dynamics.
\end{abstract}


\maketitle
The Fermi-Pasta-Ulam (FPU) recurrence was first raised by Fermi and his colleagues in the 1950s \cite{Fermi1955studies}. In a numerical simulation of string oscillation with nonlinear coupling between different modes to test thermalization theory, they found at a certain point the energy will return to the fundamentally excited mode, rather than distributing homogenously among different modes. This discovery triggered the rigorous investigation on plasma physics by Zubusky and Kruskal \cite{Zabusky_PRL1965interaction}, which led to the discovery of solitons. Solitons and their related theory have revolutionized the research in diverse arenas, including fluid dynamics \cite{Akhmediev_PRL2011Rogue}, optics \cite{Gordon_PRL1980Soliton,Dudley_NP2009Ten}, Bose-Eistein condensation \cite{Phillips_Science2000BEC,Lewenstein_PRL1999BEC}.

In optics, the FPU recurrence was first demonstrated based on the modulation instability (MI) in optical fibers \cite{Simaeys_PRL2001experimental}. As a feature of the FPU recurrence, the powers of the pump mode and the signal mode in MI evolves periodically with a phase delay of $\pi$. The collision between solitons in fibers also facilitated the observation of FPU recurrence in an active cavity \cite{Patton_PRL2007collision}. Furthermore, optical breathers, e.g., the Akhmediev breather (AB) in the nonlinear schr\"{o}dinger equation (NLSE) \cite{Akhmediev_TMP1986AB,AKhmediev_PLA2011recurrence,Dudley_NP2010peregrine}, are an important manifestation of FPU recurrence. Since collisions between breathers and solitons can result in optical rogue waves \cite{Jalali_Nature2007optical,Dudley_NP2014instabilities,Akhmediev_PRL2016integrable}, studying FPU recurrence and the control of the transition between solitons and breathers may contribute to the understanding of optical rogue waves.

Recently, maturity in the fabrication of high-Q microresonators \cite{Vahala_Nature2003optical} has fueled rapid progress on Kerr frequency comb generation \cite{Kippenberg_Nature2007optical,Gaeta_NP2010cmos,Morandotti_NP2010cmos,
Kippenberg_Science2011microresonator,Kippenberg_NP2014temporal,Weiner_NP2015mode}. In the frequency domain, microresonators based frequency comb synthesis has promising applications in optical clock \cite{Diddams_Optical2014microresonator}, optical arbitrary waveform generation \cite{Weiner_NP2011spectral}, and microwave photonics \cite{Weiner_JLT2014programmable,Mastko_NC2015high} etc. In the time domain, microresonators provide a new and important approach to realize optical solitons  \cite{Kippenberg_NP2014temporal,Kippenberg_Science2016photonic,Vahala_Optica2015soliton,Weiner_OE2016intracavity,
Gaeta_OL2016Thermal}. Different from mode-locked lasers, passive microresonators have no active gain or saturable absorber, making them free from the influence of the complex gain dynamics. Hence, soliton generation in microresonators can exhibit excellent predictability. Moreover, the bandwidth and peak power of the soliton can be controlled by varying the pump phase detuning \cite{Kippenberg_NP2014temporal,Vahala_Optica2015soliton,Bao_PRA2015CEP}. The ability to accurately predict and control soliton dynamics in microresonators accurately will make microresonators a versatile test bed for the study of fundamental soliton physics, including the FPU recurrence. Moreover, breather solitons, which can exhibit FPU recurrence, have been widely predicted in microresonators \cite{Matsko_OL2012excitation,Coen_OE2013dynamics,Bao_JosaB2014mode,Chembo_PRA2014stability,Bao_OE2015soliton,
Skryabin_PRA2015solitons}, but still lack rigorous experimental investigation, to our knowledge.

\begin{figure*}[htp]
\centering
\includegraphics[width=1.9\columnwidth]{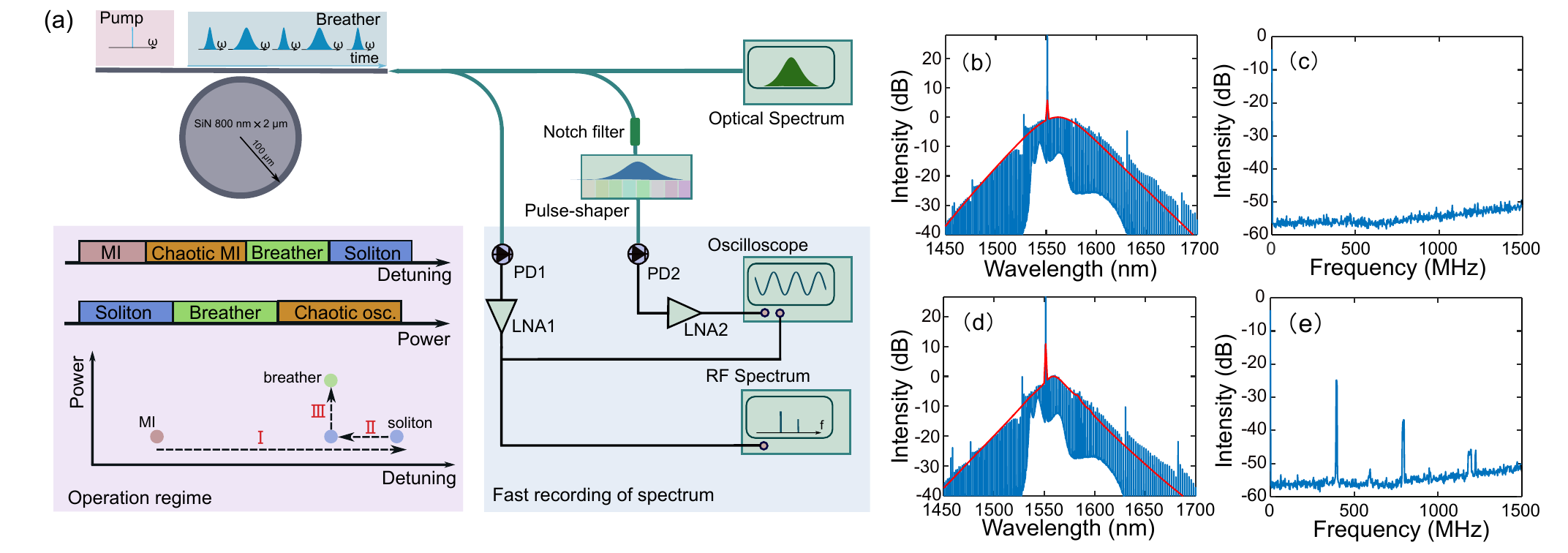}
\caption{(color online). (a) The experimental setup and operating regimes of Kerr frequency combs. Breather solitons are generated at relatively small detuning and high pump power and we use three steps illustrated by \uppercase\expandafter{\romannumeral1}, \uppercase\expandafter{\romannumeral2}, \uppercase\expandafter{\romannumeral3} to generate the breather soliton. MI: modulation instability, PD: photodiode, LNA: low noise amplifier, osc: oscillation. (b) The optical spectrum of the stable soliton (blue) and the simulated spectrum from the generalized LLE (red line). (c) The RF spectrum of the soliton state. (d) The averaged spectrum of the breather soliton spectrum from the optical spectrum analyzer (blue) and from simulation (red line). (e) The RF spectrum of the breather soliton. The pumped resonance has a linewidth of 100 MHz.}
\label{Fig1Setup}
\end{figure*}

In this Letter, we present the observation of FPU recurrence induced by breather solitons in a siicon-nitride (SiN) microresonator. By controlling the pump condition, we can excite the breather soliton in the microresonator and show the power evolution for the comb lines at the center and the wing is out of phase. We also observe the spectral breathing is not symmetric with respect to the spectrum center and identify that stimulated Raman scattering (SRS) is responsible for the symmetry breaking. Both the recurrence and the symmetry breaking are well described by numerical simulations based on the generalized Lugitato-Lefever equation (LLE), including the Raman effect \cite{Lugiato_PRL1987spatial,Bao_OE2015soliton,Skryabin_PRA2015solitons,
Kippenebrg_PRL2016raman,Wilner_OL2014efficiency}. The observation of FPU recurrence in microresonators will improve our understanding of nonlinear systems and breather theory. Furthermore, it also gives more insight into the different operating regimes of Kerr frequency combs and adds to the understanding of soliton mode-locking in microresonators.

The experimental scheme is shown in Fig. \ref{Fig1Setup}(a). An on-chip SiN microresonator is used for the generation of the Kerr frequency comb and the breather soliton. When the microresonaor is pumped by a continuous-wave laser, the power enhancement inside the cavity initiates parametric oscillation and frequency comb generation under moderate pump power. In experiments, we use an anomalous dispersion microresonator, whose dimensions are 800$\times$2000 nm with a radius of 100 $\mu$m, to generate the Kerr frequency comb. A notch filter (1550 nm, 4 nm bandwidth) is used to suppress the strong pump line by over 35 dB. Figure \ref{Fig1Setup}(a) also illustrates the operating regimes of Kerr frequency combs and the experimental method to excite the breather soliton. The breather soliton regime is close to the soliton regime and breather solitons are generated at relatively small pump phase detuning and high pump power \cite{Coen_OE2013dynamics,Bao_PRA2015CEP}. Therefore, we have three steps to generate the breather soliton: \uppercase\expandafter{\romannumeral1}) tune the laser to generate the soliton, \uppercase\expandafter{\romannumeral2}) tune the pump laser backward several picometers to lower the excitation threshold of breather solitons (see Fig. S1 in Supplementary Information \cite{SI,Weiner2011ultrafast,Silberberg_OL1997noiselike}), \uppercase\expandafter{\romannumeral3}) increase the pump power to excite the breather soliton.

For phase \uppercase\expandafter{\romannumeral1}, stable solitons can be generated by scanning the laser across the cavity resonance from blue-side to red-side \cite{Kippenberg_NP2014temporal}. To overcome the transient instability in soliton generation in SiN microresonators, the laser is tuned backward after crossing the cavity resonance \cite{Weiner_OE2016intracavity,Kippenberg_arXiv2016universal}. This backwards tuning also gives access to single soliton. At the pump power of $\sim$ 300 mW (in the bus waveguide), when pumped around 1551.28 nm, a stable single soliton is generated in the microresonator. The comb has a well-defined sech$^2$ spectrum (Fig. \ref{Fig1Setup}(b)) and low intensity noise (Fig. \ref{Fig1Setup}(c)). There are some spectral jumps on the optical spectrum, which can be attributed to the mode-interaction, however they will not change the soliton property significantly \cite{Vahala_Optica2015soliton,Weiner_OE2016intracavity,Kippenberg_PRL2014mode}, qualitatively different from the multi-phase-step combs in Ref. \cite{Diddams_NC2015phase}.

Soliton behavior in microresonators is governed by the LLE \cite{Lugiato_PRL1987spatial,Coen_OL2013modeling}. For SiN microresonators, SRS is important in determining the property of solitons \cite{Weiner_OE2016intracavity,Kippenebrg_PRL2016raman,Bao_OE2015soliton,Wilner_OL2014efficiency}. Hence, we use the generalized LLE, with SRS included, to describe the soliton generation,

\begin{small}
\begin{equation}
\begin{aligned}
& \left( {{\tau }_{0}}\frac{\partial }{\partial t}+\frac{{{\alpha}}+\theta }{2}+i{{\delta }_{0}}+i\frac{{{\beta }_{2}}L}{2}\frac{{{\partial }^{2}}}{\partial {{\tau }^{2}}} \right)E-i(1-f_{R})\gamma L{{\left| E \right|}^{2}}E \\
& -if_{R}{{\gamma }}L\left( E\int_{-\infty }^{\tau }{{{h}_{R}}\left( \tau -\tau ' \right){{\left| E \right|}^{2}}d\tau '} \right)-\sqrt{\theta }{{E}_{in}}=0, \\
\end{aligned}
\label{EqLLE}
\end{equation}
\end{small}
where $E$ is the envelope of the intracavity field, $\tau_{0}$ is the round-trip time (4.5 ps), $L$ is the length of the cavity, $t$ and $\tau$ are the slow and fast time respectively, $\alpha$ and $\theta$ are the intrinsic loss and the external coupling coefficient respectively, $\beta_2$ is the group velocity dispersion, $\gamma$ is the nonlinear coefficient, $\delta_{0}$ is the pump phase detuning, $\left|E_{in}\right|^{2}$ is the pump power, and $h_R(\tau)$ is the Raman response function. In simulations, the Raman effect is calculated in the frequency domain, with a Lorentzian gain spectrum \cite{Lin_OE2007nonlinear}, whose peak is centered at $-$14.3 THz and bandwidth is 2.12 THz. When choosing $\alpha$=0.0024, $\theta$=0.0004, $\beta_2=-$81 ps$^{2}$/km, $\gamma$=0.9 W/m, $f_{R}$=0.13 and setting the pump condition as 220 mW, and the $\delta_{0}$=0.022, a stable soliton is generated, whose spectrum is shown by the red-line in Fig. \ref{Fig1Setup}(b), in close agreement with experiments. The difference in the amplitude of the pump line results from the strong directly transmitted pump, superimposed on the output comb. A closer agreement can be reached, if we include a small third order dispersion (TOD). However, to rule out the role of TOD in the symmetry breaking of the spectral breathing, discussed below, we exclude it in the simulation (see sections 2 and 5 in Supplementary Information).

\begin{figure}[t]
\centering
\includegraphics[width=0.95\columnwidth]{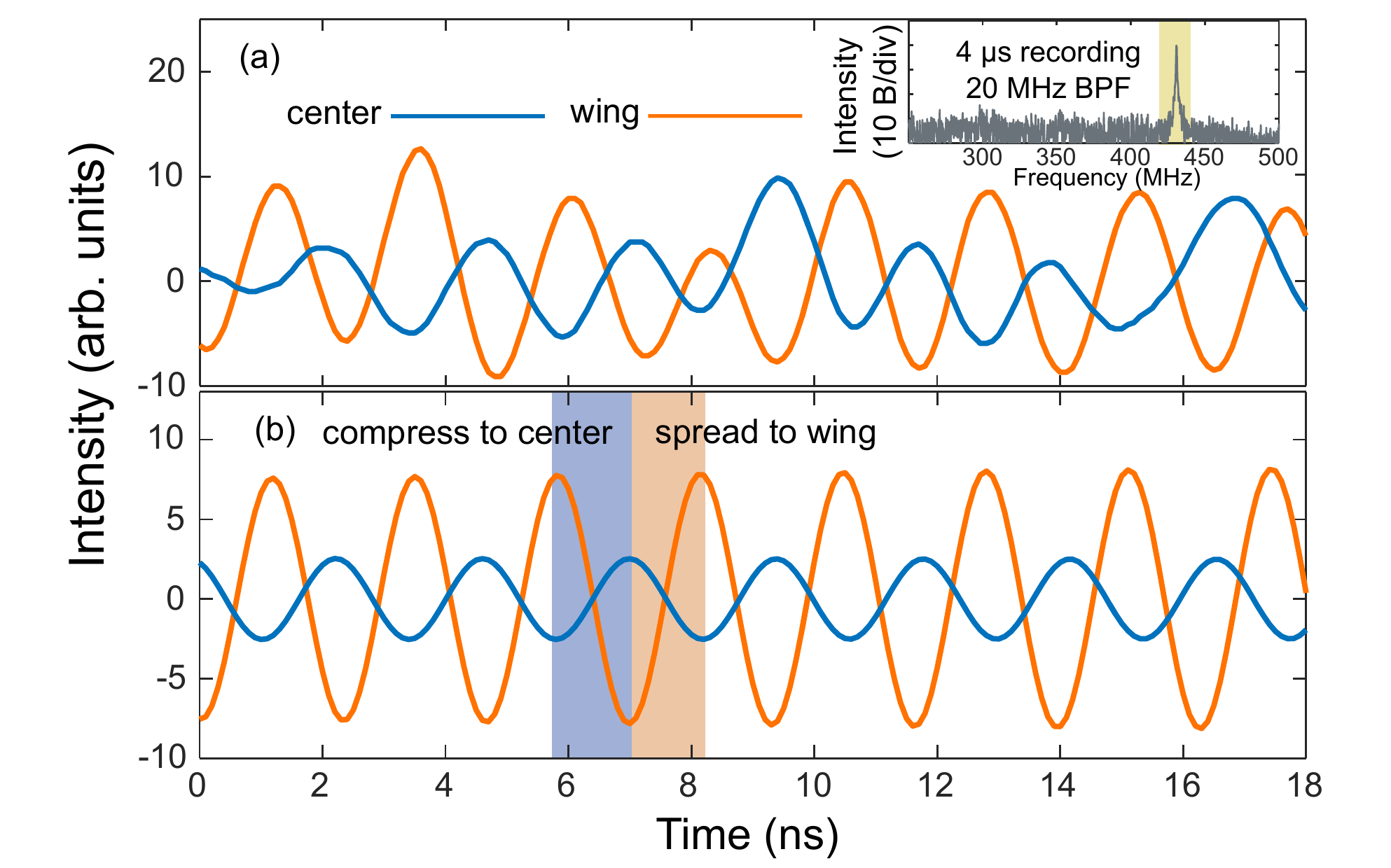}
\caption{(color online). (a) The recorded fast evolution of the comb lines around the center and in the wings of the spectrum. The inset is the RF spectrum of the power evolution measured over 4 $\mu$s for comb lines in the wings. A 20 MHz bandpass filter is used to select the strongest RF tone. (b) Reconstructed comb line power evolution after numerical filtering. The blue (orange) shaded regions illustrate the time slots where the power flows toward the center (wing). Center (blue lines): comb lines within 1553 nm$\sim$1569 nm, wing (orange lines): comb lines within 1530 nm$\sim$1550 nm and 1569 nm$\sim$1600 nm.}
\label{Fig2}
\end{figure}

\begin{figure}[t]
\centering
\includegraphics[width=0.75\columnwidth]{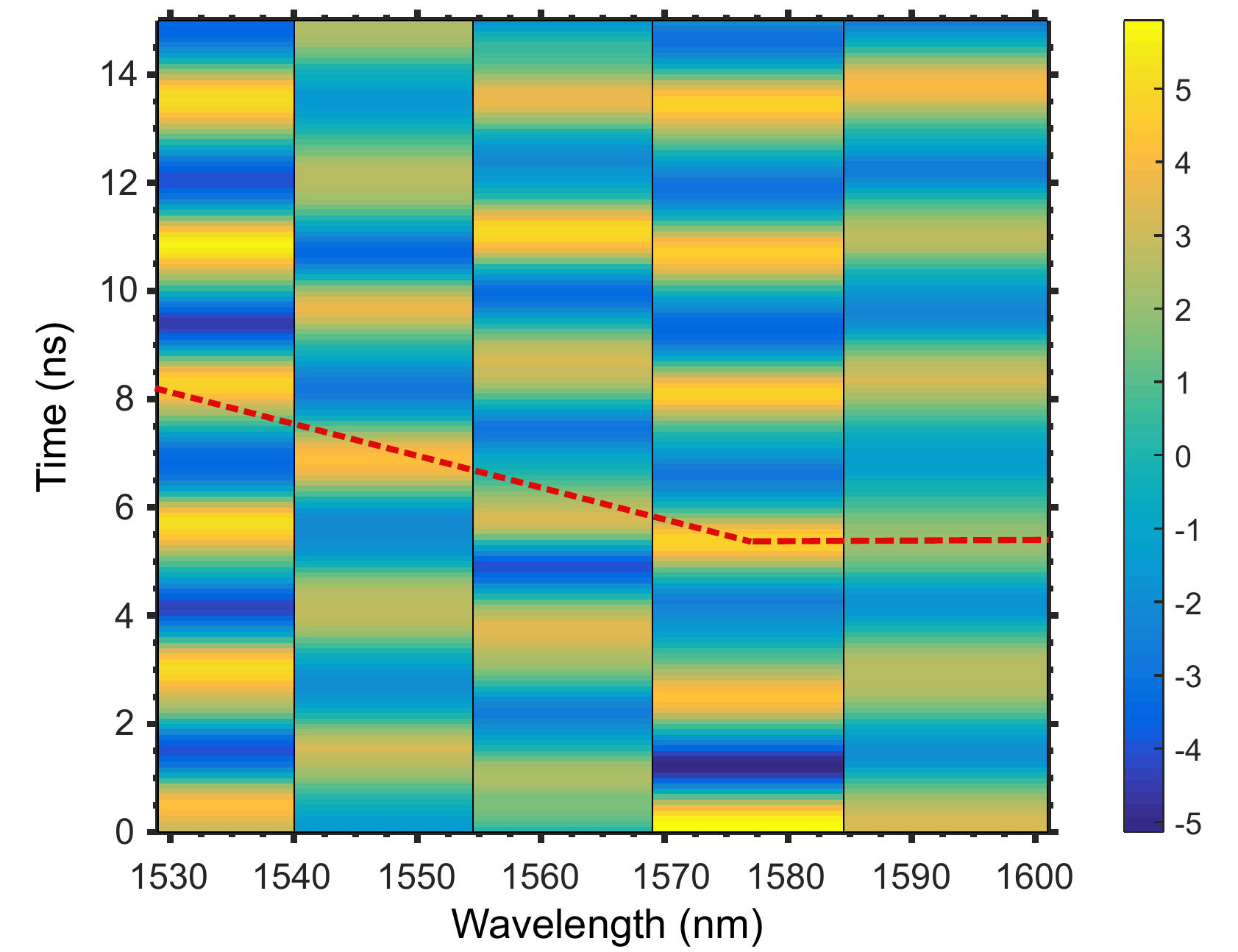}
\caption{(color online). The recorded power evolution for different comb line groups when the pulse-shaper is used to select spectral regions with finer resolution. The red dashed line illustrates the peak of different groups and shows different groups experience modulations with different relative phases.}
\label{Fig3}
\end{figure}
After the generation of stable solitons, we follow the road-map shown in Fig. 1(a) to generate the breather soliton. By tuning the laser backwards several picometers further and increasing the pump power to $\sim$430 mW, the breather soliton can be excited (see section 3 of Supplemental Information for the transition dynamics from stable solitons to breather solitons). The breather soliton state is identified by the sharp peak in the RF spectrum (Fig. \ref{Fig1Setup}(e)). Note that the breathing frequency is nearly 4 times the linewidth of the pumped resonance (100 MHz). From our measurements, the modulation depth of the converted comb lines, defined as ${\left( {{P}_{\max }}-{{P}_{\min }} \right)}/{\left( {{P}_{\max }}+{{P}_{\min }} \right)}$ with $P_{\max(\min)}$ being the maximum (minimum) average power, is $\sim$50$\%$. Similar narrow RF peaks were observed in normal dispersion microresonators and interpreted as dark breather pulse \cite{Weiner_NP2015mode}. Narrow RF peaks have also been reported recently in anomalous dispersion Si and SiN microresonators \cite{Gaeta_CLEO2016}. The spectrum of the breather solitons becomes sharper at the top of the spectrum, compared to the soliton spectrum. In simulations, breather solitons can be generated by decreasing $\delta_{0}$=0.014 and increasing the pump power to 360 mW. The averaged spectrum (averaging over slow time $t$) of the simulated breather soliton is shown by the red line in Fig. \ref{Fig1Setup}(d), which reproduces the sharp top of the experimental spectrum. The breather soliton retains a Raman induced frequency shift, implying the breather soliton remains as a pulse, as chaotic waveforms do not exhibit the frequency shift \cite{Kippenebrg_PRL2016raman}. The autocorrelation trace also provides evidence of pulse-like behavior (see section 4 of Supplementary Information for further temporal details and discussion of the temporal breathing). Furthermore, the excitation process is reversible, i.e., we can return to the soliton state from the breather soliton state by manually decreasing the pump power.

\begin{figure*}[t]

\centering
\includegraphics[width=1.88\columnwidth]{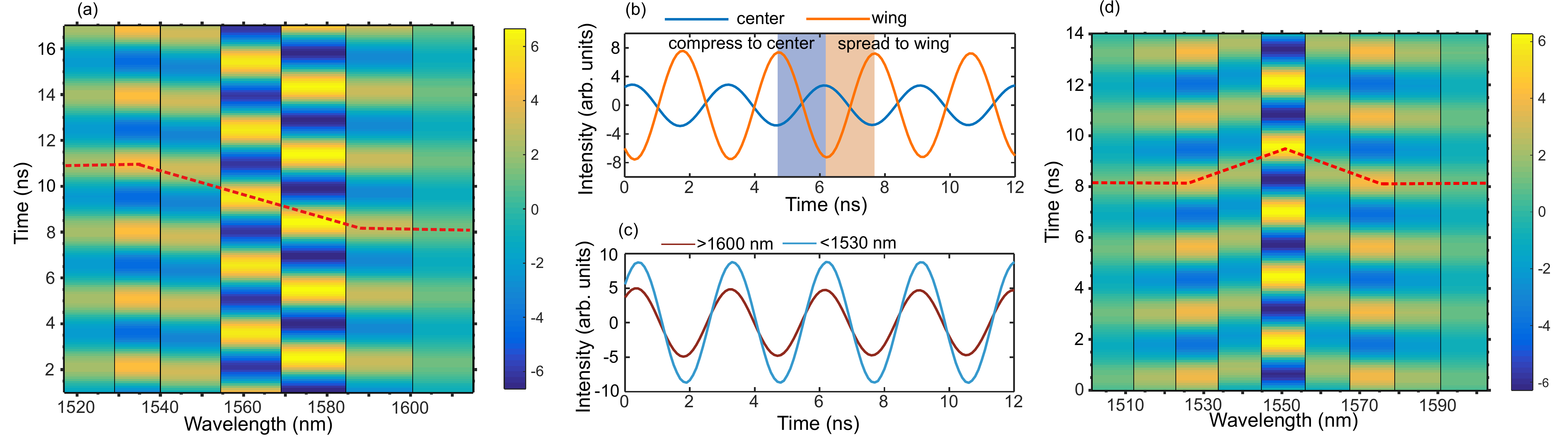}
\caption{
(color online). (a) The numerically filtered spectral evolution of the breather soliton in simulation, with SRS included. The various spectral groups far away from the center have approximately the same phase but their phase is clearly different from that of the group of comb lines around the center ($\sim$1560 nm). (b) Simulation of the FPU recurrence for comb lines around the center (blue line, 1552 nm$\sim$1567 nm) and comb lines in the wings (orange line, 1521 nm$\sim$1550 nm and 1567 nm$\sim$1592 nm, with the pump at 1551 nm excluded). The shaded regions illustrate the power exchange between the center and the wing. (c) The experimental traces at long (above 1600 nm) and short wavelengths (below 1530 nm), showing the same phase between them. (d) The numerically filtered spectral breathing in simulation, without SRS. The phase for different slices is symmetric with respect to the center of the spectrum (1551 nm). The red dashed lines in (a), (d) illustrate the peak of different slices.
}
\label{Fig4}
\end{figure*}

To test the FPU recurrence induced by breather solitons, we use a pulse-shaper \cite{Weiner_RSI2000femtosecond} to select out some specific comb lines to record the fast evolution of the breathing spectrum. The pulse-shaper has a transmission bandwidth spanning from 1530 nm to 1600 nm, with the ability to programmably select out specific comb lines. For synchronization between different spectral slices, a portion of the output comb is used as reference signal to trigger the oscilloscope (PD1 in Fig. \ref{Fig1Setup}(a)). We use the pulse-shaper to select out either 9 comb lines around the center of the spectrum ($\sim$ 1560 nm, 9 nm to the red of the pump） or the remaining 31 comb lines (but not the pump line) within the passband of the pulse-shaper in the short and long wavelength wings.

The recorded traces for the comb lines around the center and comb lines in the wings are depicted in Fig. \ref{Fig2}(a). The power change of the comb lines around the center is nearly out of phase of the comb lines in the wings, a signature of FPU recurrence. Due to the low power of a single comb line after the pulse-shaper and its fast breathing rate ($<$1 $\mu$W, $>$350 MHz), the recorded trace shows some distortions. However, the breathing exhibits good periodicity (see Fig. 1(e)), allowing numerical filtering of the signal. Here, we record the traces over 4 $\mu$s, corresponding to more than 1000 breathing cycles; the RF spectrum computed from the recorded trace for the wing is shown in the inset of Fig. 2(a). The strongest RF tone is selected by a 20 MHz numerical filter and inverse Fourier transformed to yield the reconstructed traces in Fig. 2(b), providing a clearer view of the recurrence.

To gain more insights into the breathing of the spectrum, the pulse-shaper is programmed to select different spectral regions continuously across the spectrum. The recorded map of the spectral breathing is presented in Fig. 3. The dashed line in Fig. 3 clearly illustrates the phase delay between different slices when the spectrum is breathing. Furthermore, the phase delay is not symmetric with respect to the center of the spectrum ($\sim$1560 nm).

Breather solitons in microresonators arise from the Hopf bifurcation in LLE \cite{Matsko_OL2012excitation,Coen_OE2013dynamics,Chembo_PRA2014stability,Skryabin_PRA2015solitons} and can also be attributed to the mismatch between the carrier-envelope phase slip of the soliton and the pump phase detuning \cite{Bao_PRA2015CEP}. To understand the FPU recurrence and the symmetry breaking in breather solitons, we use simulation to look into the breathing dynamics. To explicitly show the phase delay between different slices, the breather soliton evolution is numerically filtered in a similar way to that used in Fig. 2. The phase delay between different slices observed in experiment is also seen in the simulated breathing dynamics with SRS included, shown in Fig. 4(a) (this effect is also highlighted by the normalized spectral breathing dynamics in Fig. S5 in the Supplementary Information, which also shows TOD alone is insufficient to cause asymmetric breathing). Moreover, the symmetry breaking with respect to the spectrum center is also in agreement with Fig. 3. As shown in Fig. 4(b), the comb lines at the center and in the wing show out of phase evolution and FPU recurrence, similar to the experiment. However, the comb lines far away from the center (beyond the bandwidth of our commercial pulse-shaper) breathes with the phase delay of 2$\pi$, i.e., the same phase. This is verified by recording and numerical filtering the evolution recorded for comb lines at the long and short wavelengths selected using a home-built pulse-shaper, see Fig. 4(c). Furthermore, simulation also shows the mode-interaction induced spectral jumps in Fig. \ref{Fig1Setup}(d) have negligible influence on the breathing dynamics (see section 6 in the Supplementary Information).

To further unveil what causes the asymmetric breathing for the breather solitons, we turn off the SRS term in simulation. The breather soliton can still be excited under the same pump condition. However, the breathing dynamics in Fig. 4(d) are symmetric with respect to the center of the spectrum (coincide with the 1551 nm pump, as there is no SRS induced soliton frequency shift). The comparison with Fig. 4(a) reveals that SRS is responsible for the symmetry breaking in the breathing dynamics observed in Fig. 3. SRS is generally significant for SiN microresonators \cite{Kippenebrg_PRL2016raman,Weiner_OE2016intracavity}. However, in fluoride microresonators, SRS is much weaker \cite{Kippenberg_NP2014temporal}; hence, symmetric breathing of solitons can be expected.

For the breather in the microresonator, one significant difference from ABs is the energy in the wing of the spectrum can return to a group of several modes around the center (see Fig. 2 and Fig. 4(b)) while energy returns to the single pump for ABs. This is because the breather in microresonators remains to be a pulse during evolution, while ABs fully recover to be a continuous wave in FPU recurrence. This difference shows how the soliton dynamics in the framework of the LLE are distinct from those in the NLSE and illustrates how the dissipative effects and SRS break the integrability of the system and affect the breather behavior.


In conclusion, we have observed breather solitons and FPU recurrence in an on-chip SiN microresonator. Breather solitons can be excited at high pump power and small detuning. By selecting out two groups of comb lines around the center and in the wing, we find the energy returns to the center and flows out from the center periodically. 
Furthermore, we show SRS breaks the symmetry of the spectral breathing. Our results shows how the dissipative effects and SRS affect the breather properties and can contribute to the understanding of breathers and the operation of Kerr frequency combs. Furthermore, the observation of FPU recurrence in microresonators shows on-chip microresonators can be used as a powerful platform to explore soliton physics.

\begin{acknowledgments}
This work was supported in part by the Air Force Office of Scientific Research (AFOSR) (FA9550-15-1-0211), by the DARPA PULSE program (W31P40-13-1-0018) from AMRDEC, and by the National Science Foundation (NSF) (ECCS-1509578).
\end{acknowledgments}



\bibliography{reflist}
\end{document}



\title{Supplementary Information for ``Observation of Fermi-Pasta-Ulam Recurrence Induced by Breather Solitons in an Optical Microresonator"}

\author{Chengying Bao$^1$}
\email{bao33@purdue.edu}

\author{Jose A. Jaramillo-Villegas$^{1,2}$}

\author{Yi Xuan$^{1,3}$}

\author{Daniel E. Leaird$^{1}$}

\author{Minghao Qi$^{1,3}$}

\author{Andrew M. Weiner$^{1,3,4}$}
\email{amw@purdue.edu}
\affiliation{$^{1}$School of Electrical and Computer Engineering, Purdue University, 465 Northwestern Avenue, West Lafayette, IN 47907-2035, USA}
\affiliation{$^{2}$Facultad de Ingen\'{i}eras, Universidad Tecnol\'{o}gica de Pereira, Pereira, RI 66003, Colombia}
\affiliation{$^{3}$Birck Nanotechnology Center, Purdue University, 1205 West State Street, West Lafayette, Indiana 47907, USA}
\affiliation{$^{4}$Purdue Quantum Center, Purdue University, 1205 West State Street, West Lafayette, IN 47907, USA}

\maketitle

\renewcommand{\thefigure}{S\arabic{figure}}
\renewcommand{\theequation}{S\arabic{equation}}

\renewcommand*{\citenumfont}[1]{S#1}
\renewcommand*{\bibnumfmt}[1]{[S#1]}
\renewcommand{\thesection}{\arabic{section}}
\section{Excitation threshold of breather solitons at different detunings}

\begin{figure}[h]
\includegraphics[width=0.55\columnwidth]{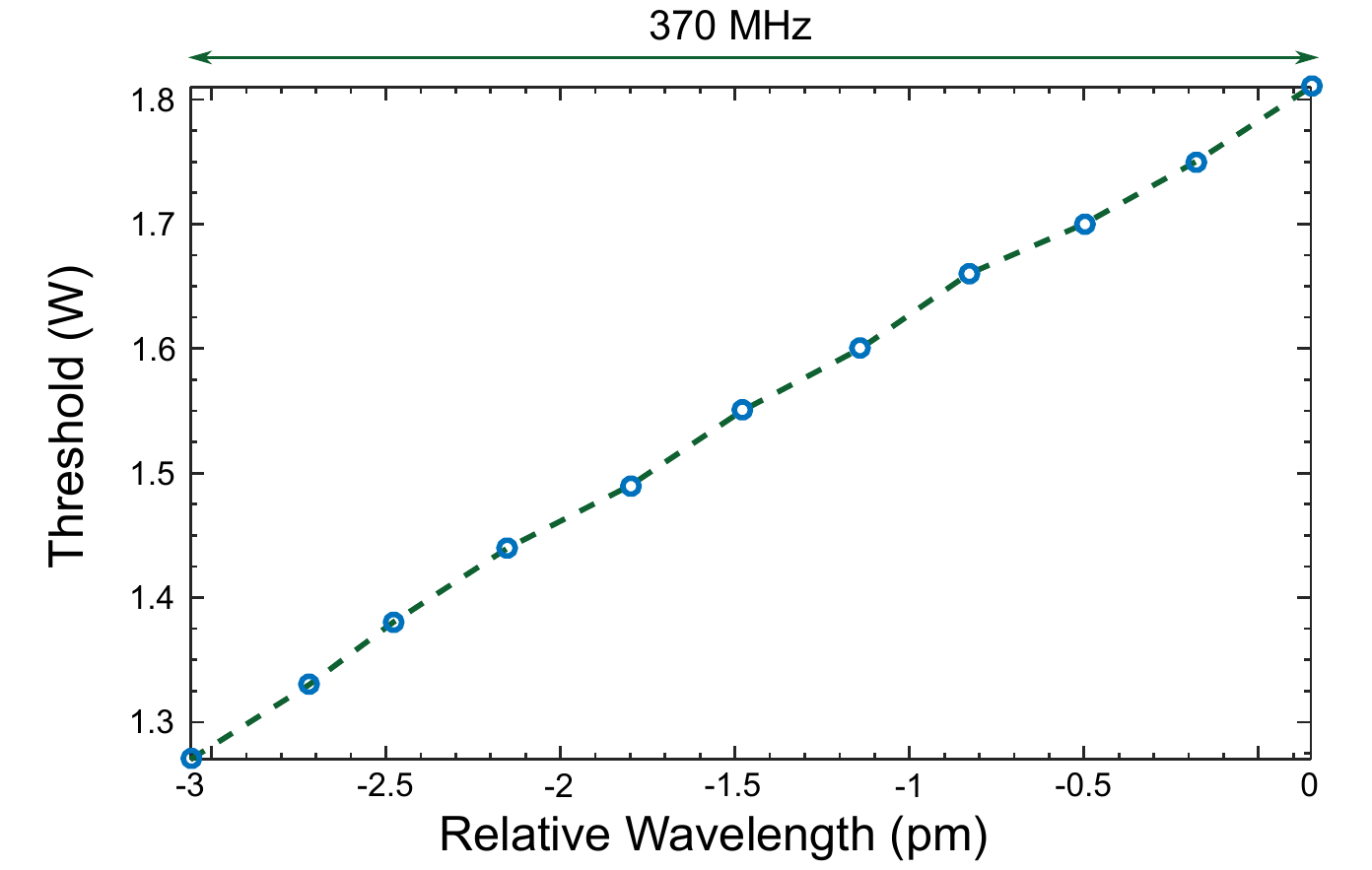}
\caption{(color online) The dependence of excitation threshold for breather solitons on final pump wavelength (detuning). The excitation threshold increases with increasing pump wavelength (increasing detuning). The threshold is recorded as the power in the fiber, prior to the waveguide.}
\label{Fig1Threshold}
\end{figure}

The excitation threshold of breather solitons increases with the pump wavelength (detuning). To measure this relationship, we first excite the stable soliton in the cavity. Then we tune the pump wavelength to different wavelengths while maintaining the soliton. Finally, we increase the pump power to excite the breather solitons at these different wavelengths and record the corresponding threshold. The threshold is shown to increase with increasing pump wavelength (detuning) in Fig. \ref{Fig1Threshold}. Note the actual power needed to generate the breather soliton depends on the fiber-to-chip coupling and polarization state.

\section{Effect of third order dispersion on the optical spectrum}

\begin{figure}[h]
\includegraphics[width=0.63\columnwidth]{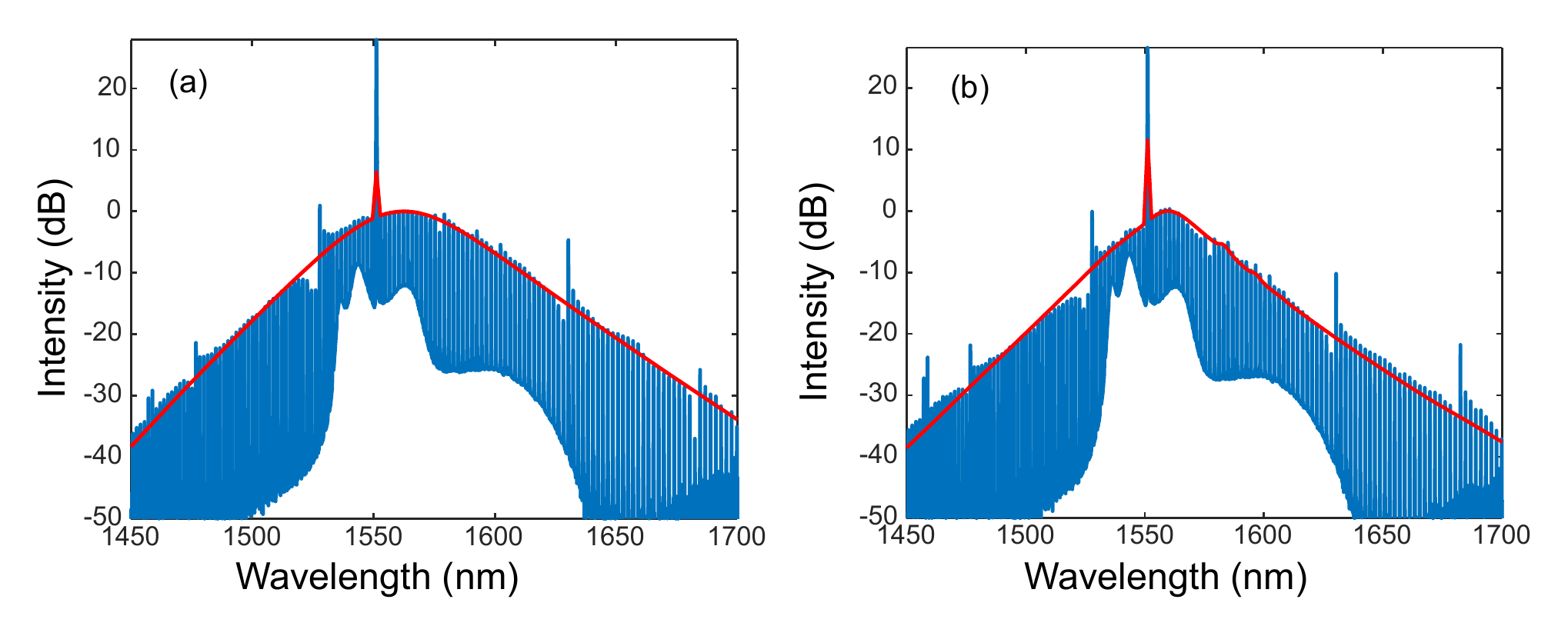}
\caption{(color online) Experimental (blue) and simulated optical spectra (red) for (a) soliton and (b) breather soliton with TOD included.}
\label{Fig2TODrole}
\end{figure}

The inclusion of third order dispersion (TOD) helps to make the agreement with the measured optical spectra even better. For example, if a TOD of $\beta_{3}=-$0.84~ps$^3$/km included in the simulation, while keeping the other parameters the same, the agreement with the experimental spectrum is improved, as shown in Fig. \ref{Fig2TODrole} especially on the long wavelength side.

\section{Generation dynamics of breather solitons}

\begin{figure}[h]
\includegraphics[width=0.8\columnwidth]{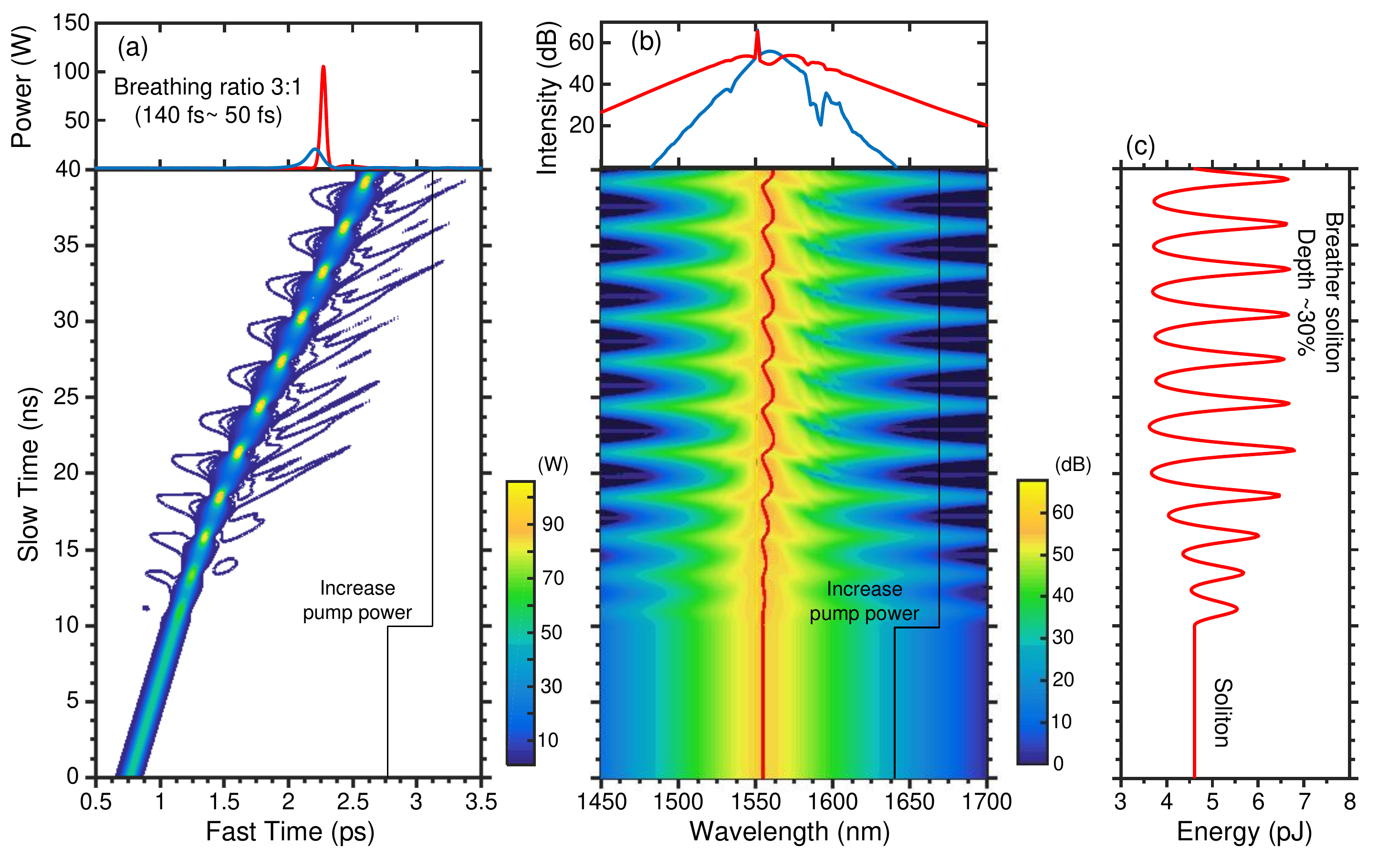}
\caption{(color online) Simulations of breather soliton dynamics. (a) The temporal dynamics of breather soliton excitation from the soliton state, in response to an abrupt increase in the pump power. Power below 1 W is represented by white color. The top panel shows the shortest and the longest pulse. (b) The spectral dynamics of breather soliton excitation. The red line shows the evolution of the center wavelength. The top panel shows the narrowest and the broadest spectra during breathing. (c) Change of the pulse energy, excluding the cw background, during the excitation of the breather soliton. After the excitation of the breather soliton, the modulation depth is about 30$\%$.}
\label{Fig5Transition}
\end{figure}
We show the temporal and spectral transition dynamics from stable soliton to breather solitons in simulation when increasing the pump power in Fig. \ref{Fig5Transition}. Here, the soliton propagates stably with 220 mW pump power and $\delta_{0}$=0.014 for times 0$\sim$10 ns. When the pump power is increased abruptly to 360 mW, the soliton starts to evolve towards the breather soliton. In the time domain, the pulse starts to experience stretching and compression. When the breather soliton is excited the temporal breathing ratio of the pulsewidth is nearly 3 (from 50 fs to 140 fs, top panel of Fig. \ref{Fig5Transition}(a)). Also, the leading and trailing tails of the pulse breathe in different manners. The optical spectrum also starts to breathe after increasing the pump power. The breathing ratio of the optical bandwidth is close to the time domain breathing ratio. The center frequency is calculated as

\begin{equation}
{{\nu }_{center}}=\frac{\int_{0}^{+\infty }{\nu {{\left| \tilde{E}(\nu ) \right|}^{2}}d\nu }}{\int_{0}^{+\infty }{{{\left| \tilde{E}(\nu ) \right|}^{2}}d\nu }},
\label{cf}
\end{equation}
where ${\left| \tilde{E}(\nu ) \right|}^{2}$ is the power spectrum. The calculated center wavelength (the strong pump line is excluded), shown by the red line in Fig. \ref{Fig5Transition}(b), is found to oscillate during breathing. However, the center wavelength is always at the red-side of the pump. The top panel of Fig. \ref{Fig5Transition}(b) shows the narrowest spectrum and the broadest spectrum during breathing. It also shows that power at the center decreases when power at the wing increases, which is a clear illustration of the FPU recurrence.

Furthermore, the intracavity pulse energy, excluding the cw background, also starts to oscillate after increasing the pump power, shown in Fig. \ref{Fig5Transition}(c). When fully developed, the modulation depth of the energy breathing is about 30$\%$. Figure \ref{Fig5Transition}(c) also shows it takes about 10 ns to accomplish the transition from soliton to breather soliton.

\section{Time domain characterization of breather solitons}

Breather solitons experience compression and stretching in the time domain (c.f. Fig. \ref{Fig5Transition}). The fast oscillation of the breather solitons makes it hard to capture their temporal breathing. However, slow (time averaged) second harmonic generation (SHG) based measurements still give some insight into the temporal breathing of the breather solitons.

\begin{figure}[h]
\includegraphics[width=0.8\columnwidth]{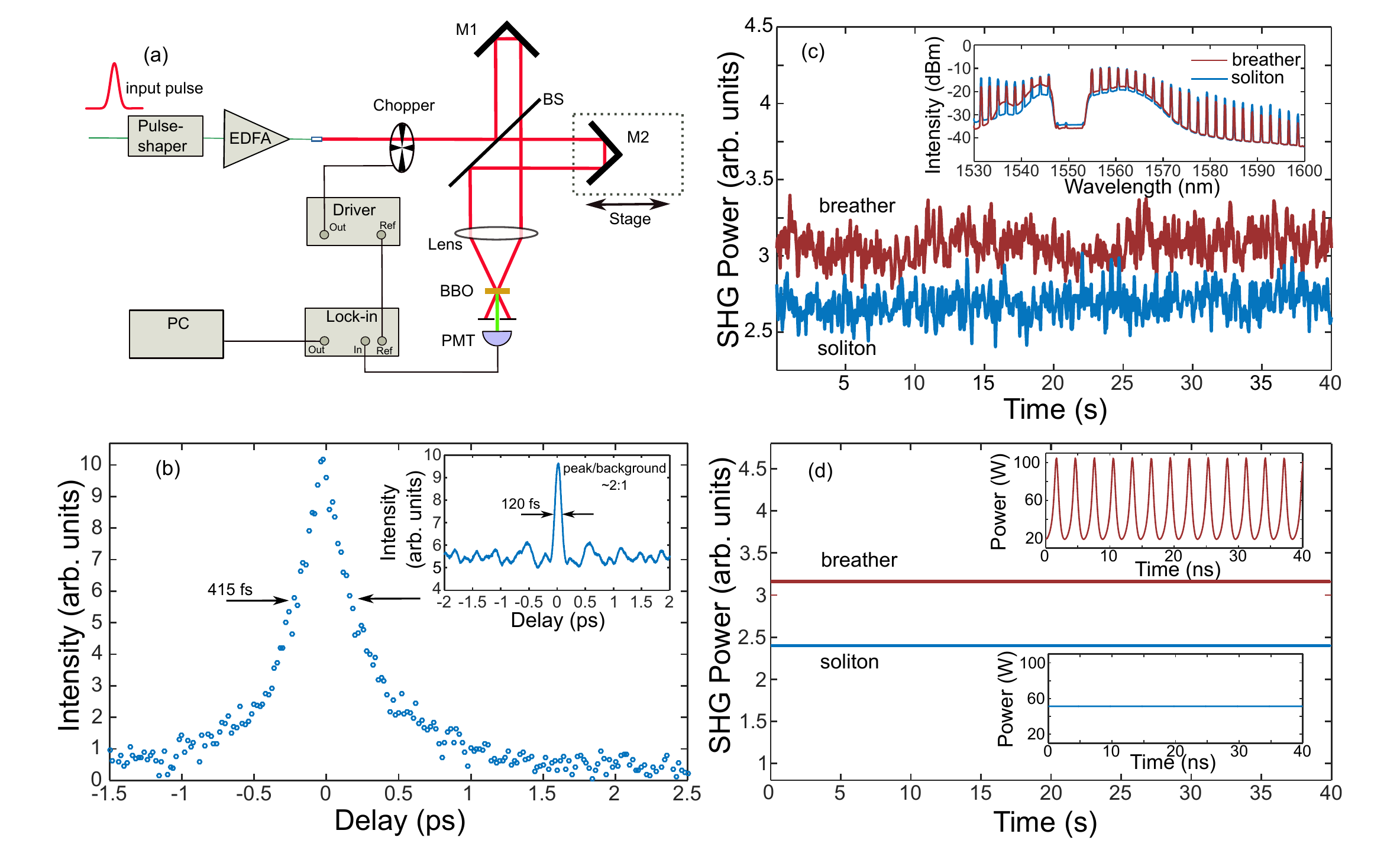}
\caption{(color online) (a) Experimental setup for autocorrelation and second harmonic generation (SHG) measurements, EDFA: erbium-doped fiber amplifier, BS: beam splitter, BBO: barium borate, PMT: photomultiplier tube, Lock-in: lock-in amplifier, PC: computer, M1, M2: mirror. (b) The measured autocorrelation trace for the breather soliton, showing pulse-like operation. The inset is the autocorrelation trace for the chaotic state, which is narrower. The peak to background ratio is about 2:1, a feature of broadband intensity noise. (c) The SHG strength for the breather soliton (red) and soliton (blue) state, at zero delay. The breather solitons produce a stronger SHG signal than do the solitons. The insets show the optical spectrum of breather and soliton after the EDFA, showing the soliton has higher average power. (d) The estimated SHG power based on simulations, demonstrating a good agreement with the experimental measurement. The insets show the peak power evolutions of the breather soliton and of the soliton. The highest peak power of breather soliton is twice as large as the soliton peak power. Both the soliton and the breather soliton in simulation are generated at the detuning of $\delta_{0}$=0.014.}
\label{Fig3SHG}
\end{figure}

The time domain characterization uses a noncollinear SHG type autocorrelator, as shown in Fig. \ref{Fig3SHG}(a). A pulse-shaper is used to suppress the residual pump entirely before amplification. It also compensates roughly the dispersion of the fiber links. To avoid the nonlinearity in the fiber links, the output of the EDFA is set at a low level ($\sim$0 dBm). The value of the autocorrelation falls to nearly zero as the delay is increased; this suggests that the breather soliton maintains a single pulse character during evolution. The autocorrelation shows a pulse width of 270 fs (assuming a sech$^{2}$ pulse deconvolution factor); note however that the precise value should be used with caution, since the experimental trace actually represents a kind of average over the different intensity profiles that occur during the breathing. Since the dispersion of the fiber link is only approximately compensated and only a portion of the spectrum is amplified for the autocorrelation, the pulsewidth is longer than the simulated values in Fig. \ref{Fig5Transition}(a). In contrast, the autocorrelation trace of the chaotic state, shown in the inset of Fig. \ref{Fig3SHG}(b), has a narrower peak. Importantly, the peak to background ratio is nearly 2:1, quite different than what we observe in the breather soliton state.  In autocorrelation the contrast between peak and background is the key feature that distinguishes between isolated short pulses and continuous intensity noise.  The 2:1 contrast that we obtain in the chaotic state is what is expected for a Gaussian random field, which corresponds to continuous intensity noise with negative exponential intensity distribution \cite{Weiner2011ultrafast,Silberberg_OL1997noiselike}. The distinct characters of the intensity autocorrelations shown in Fig. S4(b) constitute important evidence that the breather solitons remain as isolated short pulses.

When the autocorrelator is set at zero delay, the SHG output signal gives information on the peak power of the input pulse, as SHG is an intensity dependent process. The recorded SHG power from the lock-in amplifier shows that the breather soliton has a higher second-harmonic power than the soliton (Fig. \ref{Fig3SHG}(c)). Note that the soliton and the breather soliton are generated at the same pump wavelength. By integrating the amplified optical spectra (see the inset of Fig. \ref{Fig3SHG}(c)), we estimate the average power for the breather soliton and the soliton are 1 mW and 1.2 mW respectively. Although the breather soliton has slightly lower average power, it produces slightly higher second-harmonic signal.  This can be attributed to the fact that during the breathing cycle, the breather soliton can reach higher peak powers (shorter pulses) than the soliton. The measurement is also consistent with the simulation. From the insets of Fig. \ref{Fig3SHG}(d), we can see the peak power of breather solitons changes periodically, and the maximum peak power for the breather soliton is twice the peak power of solitons. We further estimate the SHG power based on the following relationship,

\begin{equation}
{{P}_{2\omega }}\propto \left\langle \int_{n\tau_{0}}^{{{n\tau }_{0}+\tau_{0}}}{{{P}_{\omega }}{{(\tau)}^{2}}d\tau} \right\rangle,
\label{SHG}
\end{equation}
where $P_{2\omega}$ is the second-harmonic power, $\tau_{0}$ is the round-trip time, $P_{\omega}(\tau)$ is the power of the intracavity waveform (breather or soliton), and $\langle~\rangle$ means averaging over round-trip number $n$. The estimated second harmonic signal for the breather soliton is higher than for the soliton (Fig. \ref{Fig3SHG}(d)). Furthermore, the ratio between the second-harmonic signal under two states is also close to the measurement. This zero-delay SHG measurement provides evidence that the breather soliton breathes not only in the frequency domain but also in the time domain and can reach higher peak power during its evolution than the soliton.

\section{Normalized spectrum evolution in simulations}
\begin{figure}[h]
\includegraphics[width=0.78\columnwidth]{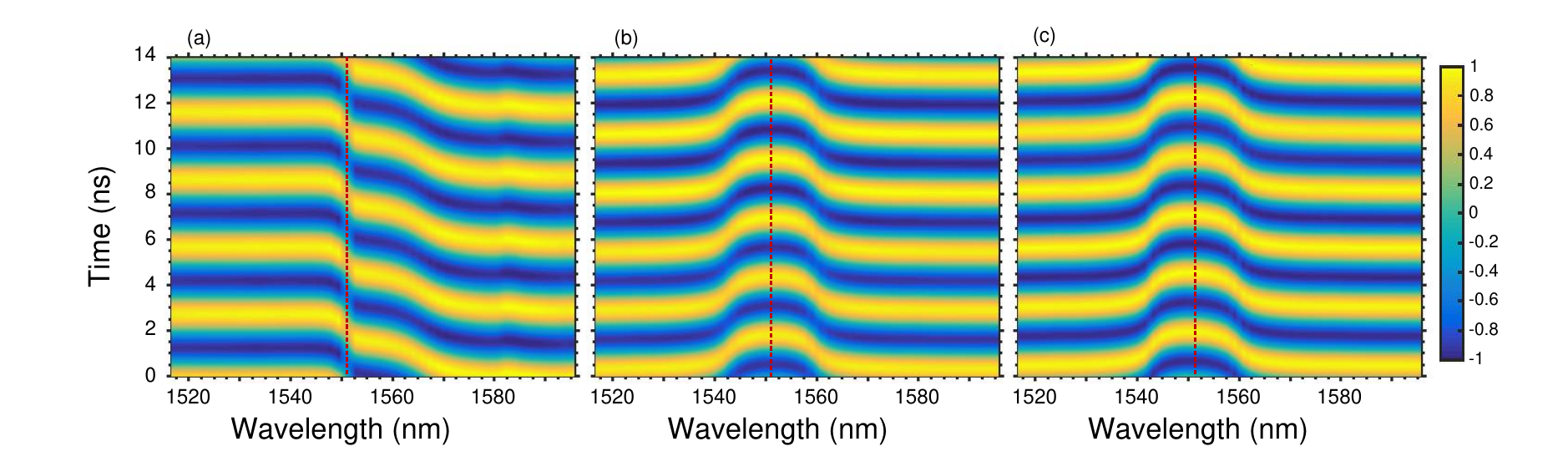}
\caption{(color online) The simulated normalized spectral breathing (a) with Raman effect included, no TOD, (b) without Raman effect, no TOD, (c) with TOD, no Raman effect. Different from Fig. \ref{Fig5Transition}, the non-oscillating component (DC component) of the spectrum is not shown in the breathing dynamics. The red lines illustrate the pump wavelength.}
\label{Fig4Symmetry}
\end{figure}

The normalized spectral evolution under different conditions is shown in Fig. \ref{Fig4Symmetry}. Unlike the main manuscript, here we show the breathing of single comb lines rather than groups of comb lines. The breathing is normalized on a line-by-line basis so that the breathing amplitude runs from $-$1 to 1. Note that the pump line is not included in Fig. \ref{Fig4Symmetry}, corresponding to filtering the pump line in experiments. In Fig. \ref{Fig4Symmetry}(a), we can see the asymmetric breathing for the Raman effect included case. When we turn off the Raman term, the breathing becomes symmetric (Fig. \ref{Fig4Symmetry}(b)). Moreover, in the case of Raman effect excluded, if we include TOD ($-$0.84~ps$^{3}$/km), the breathing is still symmetric (Fig. \ref{Fig4Symmetry}(c)). This shows that weak TOD alone is not sufficient to trigger the switch between symmetric and asymmetric breathing.

\section{Influence of mode-interaction on soliton breathing}

In simulation, we also find the mode-interaction induced spectral jumps in Fig. 1(b) have minor effect on the breathing dynamics. To include the influence of mode-interaction, we use a similar method used in \cite{Kippenberg_PRL2014mode}. We have a abrupt change on the wave-vector mismatch ($\Delta \beta^{\omega}$).

\begin{equation}
\Delta \beta ^{\omega} =\beta _{0}^{\omega }-{{\beta }_{1}}\left( \omega -{{\omega }_{0}} \right)-\beta _{0}^{{{\omega }_{0}}}=\frac{{{\beta }_{2}}}{2}{{\left( \omega -{{\omega }_{0}} \right)}^{2}}-\frac{{{a}_{1}}}{(\omega-\omega_0) -{{b}_{1}}},
\label{MI}
\end{equation}
where $\beta_{0}^{\omega}$ is the propagation constant at the frequency $\omega$, $\beta_1$ is the first-order derivative of $\beta_{0}^{\omega}$ at the frequency $\omega_{0}$ (2$\pi\times$193.4 THz). Here, we choose $a_1$=25.1 THz/m, $b_1$=24.5 THz to have an abrupt change of $\Delta \beta^{\omega}$ near the frequency $\omega_0+b_1$,

\begin{figure}[h]
\includegraphics[width=0.55\columnwidth]{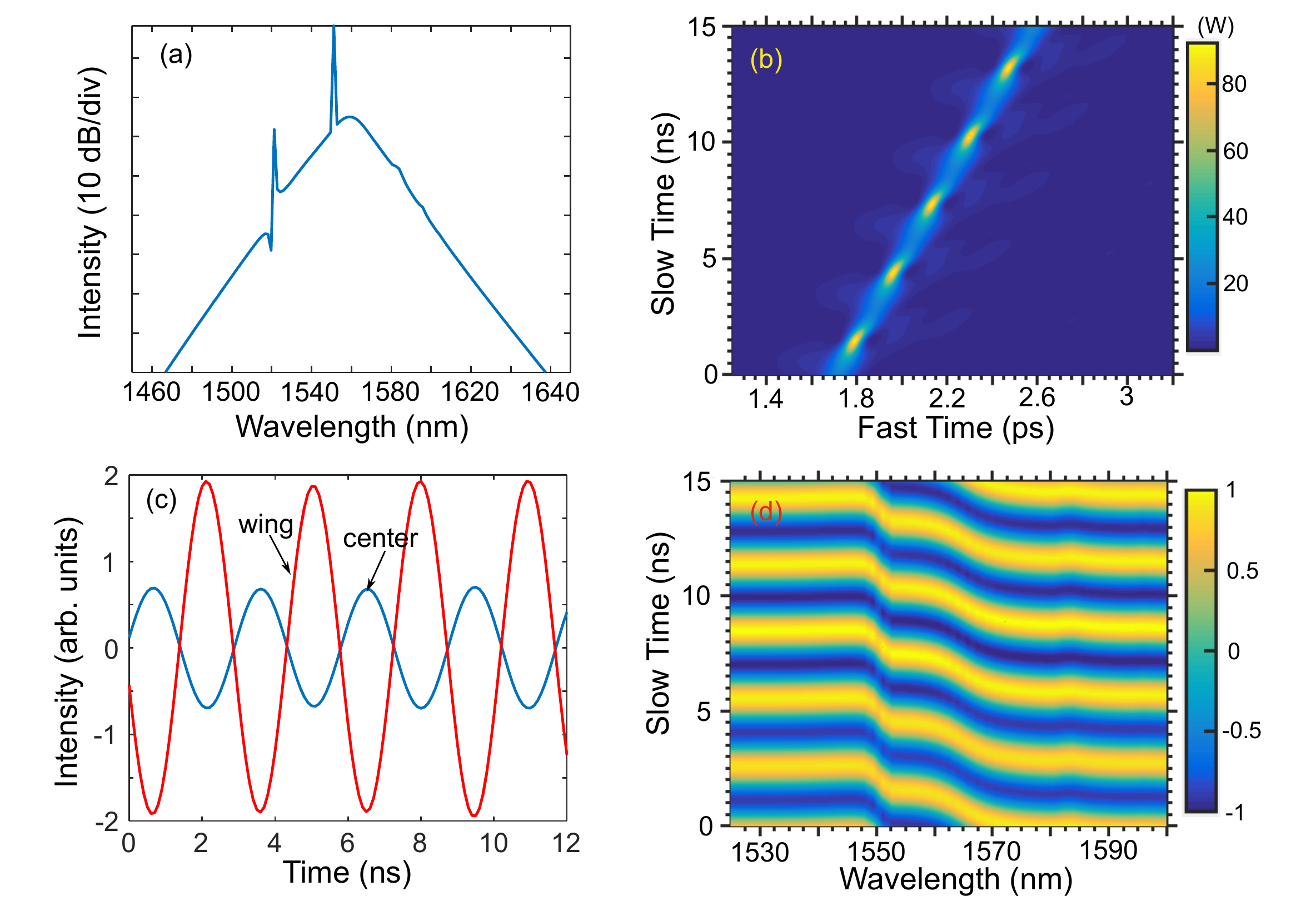}
\caption{(color online)(a) The averaged spectrum of the breather soliton, considering the mode-interaction induced spectrum spike. (b) The temporal breathing dynamics of the breather soliton under the influence of the mode-interaction. (c) The recurrence of the breather soliton, between the spectral wing (red-line) and the spectral center (blue-line). (d) The normalized spectral breathing of the breather soliton, showing asymmetric breathing with respect to the spectral center ($\sim$1560 nm).}
\label{Fig6MI}
\end{figure}
The breather soliton is still excited, using the same parameter in the manuscript. Due to the inclusion of the change on $\Delta \beta^{\omega}$, the simulated spectrum shows the spectral jumps in the averaged spectrum of the breather soliton in Fig. \ref{Fig6MI}(a), similar to the experimental measurement. Fig. \ref{Fig6MI}(b) shows the temporal breathing of the soliton. In addition, the breather soliton also shows the FPU recurrence (Fig. \ref{Fig6MI}(c)) and the asymmetric breathing (Fig. \ref{Fig6MI}(d)). Hence, we think a weak mode-interaction induced spectral jump will not change the breathing dynamics significantly.

\bibliography{reflist_SI}
\parskip 12pt